\documentclass[12pt]{article}
\usepackage{vmargin}
\setpapersize{A4}
\setmarginsrb{2cm}{3cm}{2cm}{2cm}{0pt}{0pt}{12pt}{24pt}
\usepackage[bf,sf,medium]{titlesec}
\usepackage{cite} 
\usepackage{graphicx}
\usepackage[labelfont=bf]{caption}
\usepackage{pifont}
\usepackage{authblk}
\usepackage{enumitem}
\usepackage[labelformat=simple]{subcaption}
\usepackage{floatpag}
\usepackage{amsmath}
\usepackage{amssymb}

\usepackage{relsize}
\usepackage{floatrow}

\usepackage{chngpage}
\usepackage{longtable}
\usepackage[breaklinks]{hyperref}
\hypersetup{
	colorlinks=true,
	linkcolor=blue,
	citecolor=blue
}
\DeclareMathOperator{\Expectation}{E}

\usepackage{booktabs}
\usepackage{tabularx}

\usepackage[titletoc]{appendix}

\floatpagestyle{empty}
\usepackage{bm}
\usepackage{breakurl}
\usepackage{times}
\usepackage{setspace}
\DeclareMathOperator{\MAE}{MAE}

\begin{document}

\title{A simple noise reduction method based on nonlinear forecasting}
\author[1,2]{James P.L. Tan}
\affil[1]{Interdisciplinary Graduate School, Nanyang Technological University, Singapore}
\affil[2]{Complexity Institute, Nanyang Technological University, Singapore}
\date{}
\maketitle

\begin{abstract}
Non-parametric detrending or noise reduction methods are often employed to separate trends from noisy time series when no satisfactory models exist to fit the data. However, conventional detrending methods depend on subjective choices of detrending parameters. Here, we present a simple multivariate detrending method based on available nonlinear forecasting techniques. These are in turn based on state space reconstruction for which a strong theoretical justification exists for their use in non-parametric forecasting. The detrending method presented here is conceptually similar to Schreiber's noise reduction method using state space reconstruction. However, we show that Schreiber's method contains a minor flaw that can be overcome with forecasting. Furthermore, our detrending method contains a simple but nontrivial extension to multivariate time series.  We apply the detrending method to multivariate time series generated from the Van der Pol oscillator, the Lorenz equations, the Hindmarsh-Rose model of neuronal spiking activity, and a univariate real-life measles data set. It is demonstrated that detrending heuristics can be objectively optimized with in-sample forecasting errors that correlate well with actual detrending errors. 
\end{abstract}

For time series obtained from real-world complex systems, it is often the case that one neither has nor knows an accurate mechanistic model to fit the data. Indeed, non-parametric models are becoming increasingly favored to capture the complexities and nuances that simplified mechanistic models cannot \cite{YePNAS,DeAngelisPNAS}. In the absence of any reliable mechanistic model, it becomes necessary to resort to non-parametric detrending methods to separate noise from deterministic trends. Semantically, such an endeavor may be known as noise reduction or detrending depending on what one wants to recover from the noisy time series. Regardless, there is generally no distinction between detrending and noise reduction methods since the intermediate goal of separating noise from trend is equivalent. Conventional non-parametric methods such as Loess smoothing and kernel smoothing are problematic due to the subjective choice of a time scale over which to smoothen data. Furthermore, it is unclear if recovered trends accurately represent any dynamics inherent in the time series. This ambiguity also afflicts a more recent and popular method known as Empirical Mode Decomposition (EMD) which attempts to avoid the issue of having to subjectively choose an appropriate time scale \cite{PengPNAS}. 

Here, we adopt an approach to the problem of non-parametric regression by obtaining the trend of a time series using in-sample forecasts. By casting the detrending problem as one of forecasting, we show that unambiguous trends can be objectively recovered from noisy time series. The intuition behind this endeavor is rather straightforward; a reliable forecast one time step ahead is a projection of reconstructed dynamics from available time series. Hence, a series of reliable forecasts represents a trend that captures essential dynamics inherent in the time series. We shall call such a trend a \textit{dynamical trend}. 

\section{The dynamical trend}
Let a multivariate time series $\mathbf{Y}_t$ of dimension $n$ be fully determined from its history of past states and noise terms
\begin{align}
\mathbf{Y}_t = f(\bm{\epsilon}_t, \bm{\epsilon}_{t-1}, \dots, \mathbf{Y}_{t-1}, \mathbf{Y}_{t-2}, \dots), 
\end{align}
where $\bm{\epsilon}_t$ is a multivariate random variable of dimension $l$ at time $t$ with joint probability distribution parameterized by past states
\begin{align}
\bm{\epsilon}_t \sim F(\mathbf{Y}_{t-1}, \mathbf{Y}_{t-2}, \dots). 
\end{align}
Then the dynamical trend $\mathbf{Z}_t$ of $\mathbf{Y}_t$ is defined as
\begin{align}
\mathbf{Z}_t = f_{\text{pred}}(\bm{\epsilon}_{t-1}, \bm{\epsilon}_{t-2}, \dots, \mathbf{Y}_{t-1}, \mathbf{Y}_{t-2}, \dots),
\end{align}
such that the mean squared error between $\mathbf{Z}_t$ and $\mathbf{Y}_t$
\begin{align}
\Expectation \left[ (\mathbf{Z}_t-\mathbf{Y}_t )^2 \right]
\end{align}
is minimized. Hence, $\mathbf{Z}_t$ represents the best possible forecast of $\mathbf{Y}_t$ without future knowledge of the numbers that were sampled for $\bm{\epsilon}_t$, but with knowledge of the statistical distribution of $\bm{\epsilon}_t$. 

For this paper, we shall consider the noisy multivariate time series $\mathbf{Y}_t$ of the form
\begin{align}
\mathbf{Y}_t = \mathbf{X}_t + \bm{\epsilon}_t, 
\end{align}
where $\bm{\epsilon}_t$ is a multivariate continuous random variable of joint probability density function $p(\bm{\epsilon})$ with mean $\bm{0}$, and $\mathbf{X}_t$ is a deterministic time series
\begin{align}
\mathbf{X}_t = f(\mathbf{X}_{t-1}, \mathbf{X}_{t-2}, \dots). 
\end{align}
The mean squared error is then
\begin{align}
\int_\Omega (\mathbf{Z}_t- \mathbf{Y}_t)^2 p(\bm{\epsilon}_t) \, d\epsilon_t, 
\end{align}
where the integral is over the sample space $\Omega$ of $\bm{\epsilon}_t$. Therefore, the mean squared error is minimized when $\mathbf{Z}_t = \mathbf{X}_t$. In this case, the dynamical trend is simply the time series of the deterministic system. If the goal of noise reduction is to recover $\mathbf{X}_t$ from $\mathbf{Y}_t$, then forecasting ability is equivalent to detrending performance in such a system. By associating detrending performance with forecasting ability, a detrending method can be made to be objective by optimizing its parameters based on the ability to forecast. In this paper, we will be concerned with recovering $\mathbf{X}_t$ from $\mathbf{Y}_t$, with $\bm{\epsilon}_t$ being a Gaussian white noise vector of variance $\sigma^2$. 

\section{The detrending method}
Forecasts were conducted using a class of nonlinear forecasting techniques that derive from a method known as \textit{state space reconstruction} \cite{FarmerPRL1}. First introduced in a seminal paper by Packard et al. \cite{Packard1}, and fleshed in mathematical rigor with Taken's theorem \cite{Takens1}, state space reconstruction allows for the reconstruction of a multidimensional state space from the lags of a single state variable. In this work, univariate time series were forecast using simplex projection \cite{SugiharaNature1}. For multivariate time series, multiview embedding (MVE) was used because embeddings from different combinations of variables and lags may not be equally useful in forecasting ability with the presence of noise and limited data \cite{Sauer1, YeScience1, CasdagliPhysicaD}. Instead of relying on any particular state variable, MVE selects the best combinations of variables and lags from in-sample forecasts \cite{YeScience1}. In essence, forecasting using state space reconstruction means that each corrected point is obtained by forecasting using nearest neighbors in the reconstructed state spaces one time step before. 

Utilizing state space reconstruction for the purposes of noise reduction is not new and literature on such methods exists more than two decades ago \cite{KostelichE1}. Our noise reduction method is most conceptually similar to Schreiber's method \cite{Schreiber1}. In Schreiber's method, nearest neighbors in the reconstructed state space of a point to be corrected are averaged over to produce the corrected point. This is not ideal because the noise terms that are supposed to be averaged over were involved in determining the nearest neighbors. Consider the case where nearest neighbors are identified from a noisy time series in a small neighborhood about the point to be corrected in the reconstructed state space. Then the corrected point is relatively unchanged from the original. This necessitates an increase of neighborhood size until a reasonable correction is available which means the inclusion of nearest neighbours farther away from the original point. Such a problem can be mitigated by correcting the point using a forecast one time step ahead from nearest neighbors of the state one time step before. In this way, the noise terms to be averaged over would be independent of the terms used to identify the nearest neighbors. 

In combination with these forecasting techniques, we make use of two heuristics inspired from previous literature that can be optimized with in-sample forecasting to improve detrending performance \cite{KostelichE1}: 

1) Under a time reversal, the time series also contains information on the dynamics of the system. Hence, forecasting performance may be improved if the ability to forecast backward is as good or even better than the ability to forecast forward. This leads to three possible detrending algorithms. The first is based on forward forecasting, the second is based on backward forecasting, while the third relies on a combination of both forward and backward forecasting where the forward forecast and the backward forecast are combined with a simple average. We call these three variants the forward algorithm, the backward algorithm, and the bidirectional algorithm. 

2) As pointed out by Schreiber \cite{Schreiber1}, noise reduction from a first pass of the algorithm may not be optimal. The detrending algorithm may then be applied recursively on corrected time series to improve detrending performance. Thus, the number of times the detrending algorithm is run recursively, $r$, becomes a parameter to optimize.

In state space reconstruction, multivariate time series $\mathbf{B}_t$ in a multidimensional state space of dimension $E$ (also called the embedding dimension) can be constructed from $E-1$ lags of a univariate time series $Y_t$ i.e.
\begin{align}
\mathbf{B}_t = (Y_t, Y_{t-1}, \dots , Y_{t-E+1}).
\end{align}
In simplex projection, to obtain a forecast one time step ahead for a state vector $\mathbf{B}_{0}$, the $E+1$ nearest neighbors of $\mathbf{Y}'_{0}$ are identified and the forecast is computed from the corresponding vectors of these nearest neighbors one step ahead in time \cite{SugiharaNature1}. The computation is done by averaging with exponential weights according to the Euclidean distance to $\mathbf{B}_{0}$. Therefore, the forecast $\hat{\mathbf{B}}_1$ is given by
\begin{align}
\hat{\mathbf{B}}_1 = \sum_{nni} \mathbf{B}_{nni+1} w_{nni}, 
\end{align}
where $nni$ (short for nearest neighbor index) is the time index of one of the nearest neighbors of $\mathbf{B}_0$ and
\begin{align}
w_{nni} = h\exp[-||\mathbf{B}_{nni} - \mathbf{B}_0||/\min(d)]. 
\end{align}
Here, $h$ is a normalization constant for the weights and $\min(d)$ refers to the smallest distance between $\mathbf{B}_0$ and its nearest neighbors. If only the forecast for $Y_1$ is needed, then only the first coordinate of $\hat{\mathbf{B}}_1$ needs to be computed to obtain the forecast for $Y_1$. 

In multiview embedding (MVE), state space reconstructions with embedding dimension $E$ are done for all variable and lag combinations of a multivariate time series $\mathbf{Y}_t$ such that each combination consists of at least a variable of lag 0 \cite{YeScience1}. The top $k$ reconstructions for each coordinate of $\mathbf{Y}_t$ are then chosen based on in-sample leave-one-out cross-validation (LOOCV) forecasting performance using simplex projection. In this case, in-sample forecasting performance for different embeddings is ranked by correlation between forecasts and the noisy time series. To obtain a forecast for MVE, the nearest neighbor from each reconstruction in the top $k$ reconstructions is identified and the vectors from these nearest neighbors one step ahead in time are averaged over to produce the forecast. Following Ye and Sugihara \cite{YeScience1}, we set $k=\sqrt{m}$, where $m$ is the number of available variable and lag combinations. 

\paragraph{Forward algorithm} To obtain a corrected time series from in-sample forecasts, MVE was used for multivariate time series whereas simplex projection was used for univariate time series. Here, it should be noted that for a corrected point $\hat{Y}_1$, we also made use of $Y_1$ such that
\begin{align}
\hat{Y}_1 = \alpha Y_1 + (1-\alpha) \hat{Y}_{1+}, 
\end{align}
where $0<\alpha<1$ is a real number that we set at 0.5 for all detrending done in this paper, and $\hat{Y}_{1+}$ indicates the in-sample forecast for $t=1$ using MVE/simplex projection forward in time (+). 

\paragraph{Backward algorithm} The backward algorithm is the same as the forward algorithm, except that time series were first flipped horizontally before detrending with the forward algorithm. The corrected time series were then flipped horizontally again to give the corrected time series by the forward algorithm. The forecast is
\begin{align}
\hat{Y}_1 = \alpha Y_1 + (1-\alpha) \hat{Y}_{1-}, 
\end{align}
where $\hat{Y}_{1-}$ indicates a forecast made by the backward forecasting of MVE/simplex projection. 

\paragraph{Bidirectional algorithm} The bidirectional algorithm combines the forecast of the forward algorithm and the backward algorithm by a simple average. The forecast is
\begin{align}
\hat{Y}_1 = \alpha Y_1 + (1-\alpha) \hat{Y}_{1\pm}, 
\end{align}
where $\hat{Y}_{1\pm}=(\hat{Y}_{1+}+\hat{Y}_{1-})/2$ indicates a forecast made by the forward and backward forecasting of MVE/simplex projection. 

As alluded to before, another heuristic is to run the detrending algorithms recursively on corrected time series. Let $\hat{Y}_1^{(r)}$ be the time series corrected by the bidirectional algorithm (for example) over $r$ recursive iterations from the original coordinate time series $Y_t$ such that
\begin{align}
\hat{Y}_1^{(r)} = \alpha \hat{Y}_1^{(r-1)} + (1-\alpha) \hat{Y}_{1\pm}^{(r-1)}, 
\end{align}
where $\hat{Y}_{1\pm}^{(r-1)}$ indicates the forecast made by the forward and backward forecasting of MVE/simplex projection using the time series $\hat{Y}_1^{(r-1)}$. Hence, we define $\hat{Y}_1^{(0)}=Y_1$. The in-sample cross-validation forecasting error for $\hat{Y}^{(r-1)}_1$ which we use as an estimate of the detrending performance of $\hat{Y}^{(r)}_1$ is then the mean absolute error of $\hat{Y}_{1\pm}^{(r-1)}$ measured against $Y_t$. The in-sample cross-validation error for the other two algorithms were calculated in a similar way.

\section{Results and Discussion}
We test the detrending method on noisy time series sampled from the Van der Pol oscillator, the chaotic Lorenz equations and the chaotic Hindmarsh-Rose model. Deterministic time series $\mathbf{X}_t$ sampled from these systems were combined with additive observational noise $\bm{\epsilon}_t$ to give $\mathbf{Y}_t=\mathbf{X}_t+\bm{\epsilon}_t$, the noisy time series to detrend. The noise-reduced time series is obtained using in-sample forecasts one time step ahead. The errors from these forecasts (as calculated against the noisy time series) essentially constitute a performance measure (the mean absolute error, MAE) from leave-one-out-cross-validation (LOOCV). This cross-validation error is used as an estimate of the potential detrending performance of the corrected time series obtained by the in-sample forecasts. This allows us to objectively identify the detrending parameters, i.e. which algorithm to run (forward, backward, or bidirectional) and how many times to run it recursively, based on the the lowest MAE. Ideally, the goals of detrending and forecasting are equivalent in these systems. However, we should not expect a perfect correlation between in-sample forecasting errors and actual detrending errors because in-sample forecasting errors are calculated against noisy time series whereas actual detrending errors are calculated against the deterministic time series $\mathbf{X}_t$. A significant presence of noise also leads to complications such as an inaccurate reconstruction of state space which would significantly limit the ability to recover any meaningful trend in the noisy time series. 

\begin{figure*}[p]
\floatbox[{\capbeside\thisfloatsetup{capbesideposition={left,top},capbesidewidth=6.6cm}}]{figure}[\FBwidth]
{\caption{\textbf{Performance of the detrending method} Error bars are estimates of standard errors. (\textbf{A}) The phase portrait of the noisy Van der Pol oscillator. (\textbf{B}) The phase portrait of the noise-reduced Van der Pol oscillator in red, including the deterministic Van der Pol oscillator in black. The corrected time series were calculated with the bidirectional algorithm for five and four recursive iterations for the $x$ and $y$-coordinate respectively. These optimized parameters (for the $x$-coordinate) were determined with (\textbf{C}), the in-sample cross-validation forecasting error for the $x$-coordinate versus the recursive iteration number $r$. An in-sample forecasting error for a time series cleaned $r-1$ times is associated with the potential detrending performance of the time series cleaned $r$ times. Therefore, a data point at $r$ is the in-sample forecasting error of a time series that has been cleaned $r-1$ times. (\textbf{D}) The actual detrending errors (MAE) of the corrected time series as calculated from $\mathbf{X}_t$. Also indicated on the plot is $\MAE(\epsilon_t)$, the MAE of the noisy time series as calculated from $\mathbf{X}_t$. (\textbf{E} to \textbf{H}) Same as (\textbf{A} to \textbf{D}) but for the $x$-coordinate of the Lorenz system. Corrected time series were calculated with the bidirectional algorithm with four recursive iterations. Time series are shown for ($\textbf{E}$ and $\textbf{F}$) instead of phase portraits but it should be noted that detrending was conducted concurrently for all three variables of the Lorenz system. (\textbf{I} to \textbf{L}) Same as (\textbf{E} to \textbf{H}) but for the $x$-coordinate of the Hindmarsh-Rose model. Noise-reduced time series were calculated with the bidirectional algorithm with three recursive iterations. }}
{\includegraphics[scale=0.23]{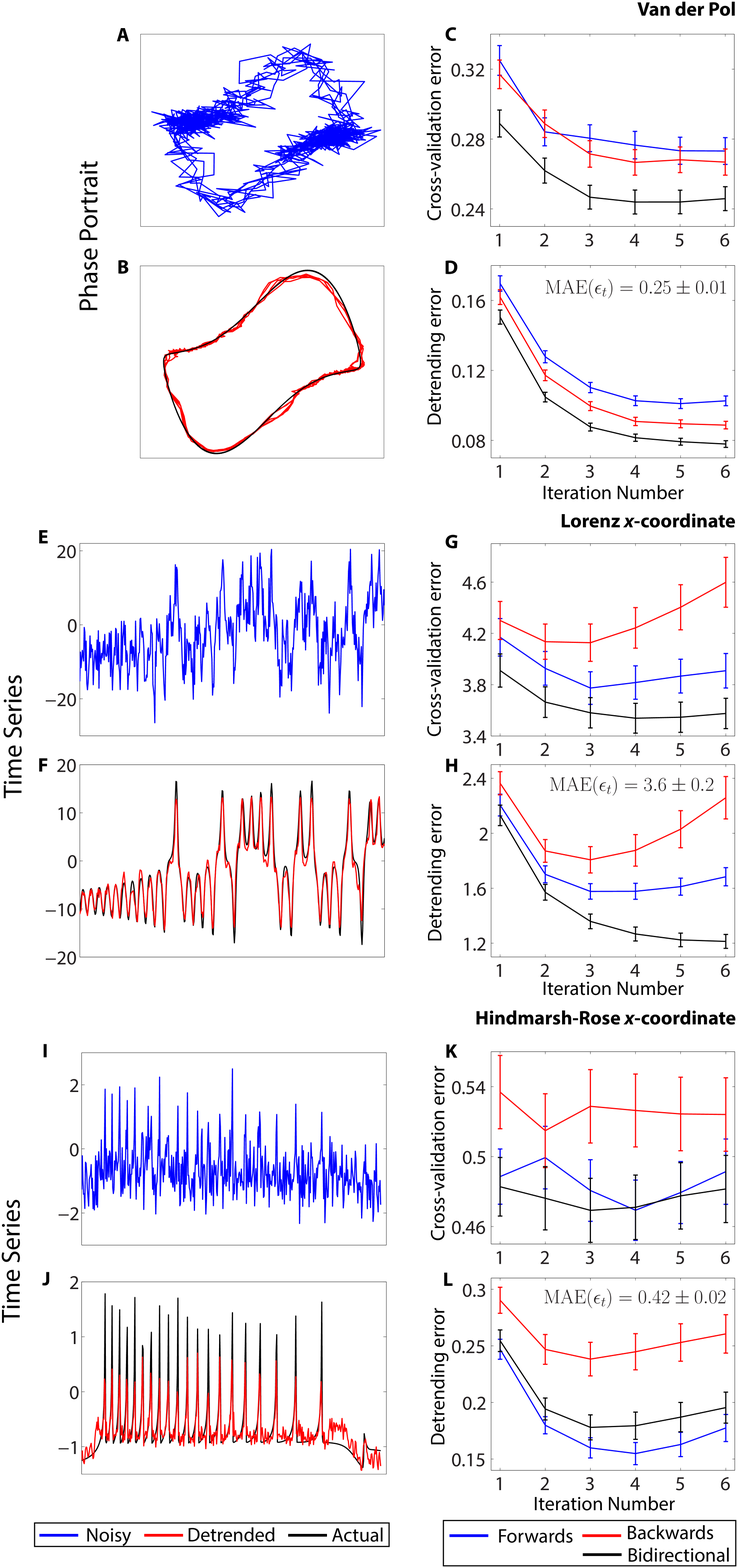}}
\label{fig:Figure1}
\end{figure*}

The results of the detrending method for several periods of the limit cycle from the Van der Pol oscillator can be seen in Fig. 1A-1D. Here, $\sigma^2=0.1$ and 800 data points were used. The in-sample forecasting errors for the $x$-coordinate (Fig. 1C) from LOOCV correlates well with the actual detrending errors (Fig. 1D) i.e. the error between the noise-reduced time series and $\mathbf{X}_t$. In particular, the in-sample errors predict that making use of the bidirectional algorithm with five recursive iterations of the algorithm would be optimal, a result that was corroborated to a good degree by the actual detrending errors. The detrending of the noisy limit cycle requires the subjective choice of the span. If $\mathbf{Y}_t$ contains only one oscillation of the limit cycle, then the behavior of the detrending algorithm presented here is conceptually similar to that of Loess smoothing in that cleaned data points are computed locally from nearest neighbors in time. This is the case because in a single oscillation, nearest neighbors in time are also nearest neighbors in space and cleaned data points in the algorithms are computed from nearest neighbors in reconstructed state spaces. However, if multiple oscillations are present, then unlike Loess smoothing, a cleaned data point can also be computed across large differences in time. In this case the detrending method confers a higher performance over Loess smoothing whatever the span (see Appendix). This is despite the fact that parameters from the detrending method were optimized objectively without knowledge of $\mathbf{X}_t$. 

\begin{figure*}[t]
\centering
\includegraphics[scale=0.45]{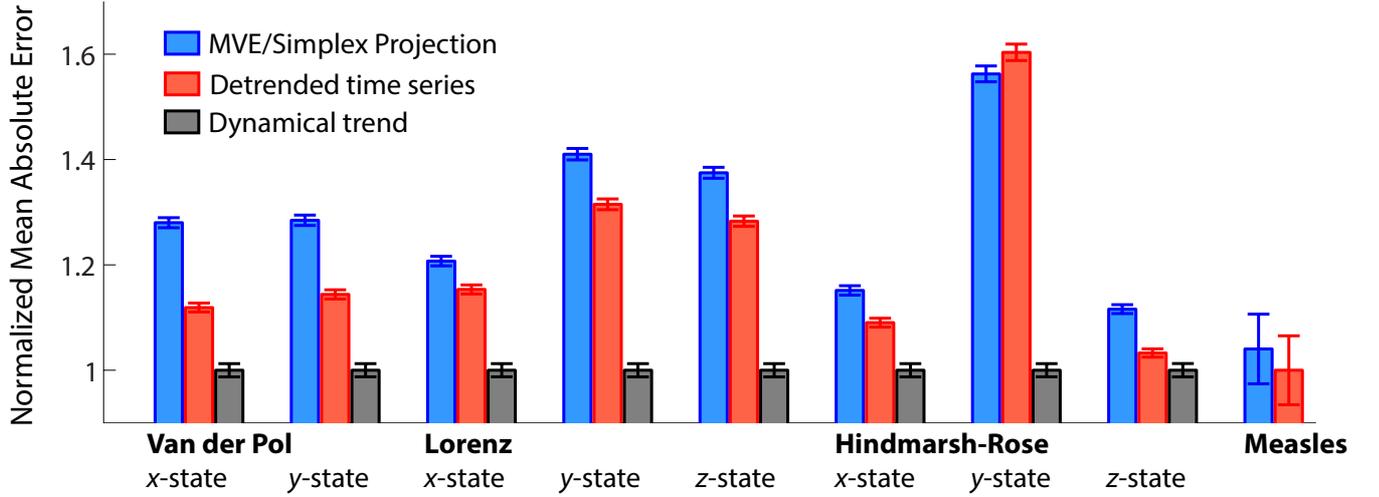}
\caption{\textbf{Out-of-sample forecast performance of noise-reduced time series} The out-of-sample forecast performance using the original noisy time series and noise-reduced time series as libraries for forecasting. Noise-reduced time series were objectively corrected with the detrending method. Forecast performance is measured by the normalized MAE for the various systems. Forecast MAEs (blue and red bars) are normalized with the MAE of the dynamical trend (black bars) which is calculated against the noisy out-of-sample time series. For the measles data set, MAEs are normalized against the MAE from the noise-reduced time series instead due to an unknown dynamical trend. Error bars are estimates of standard errors. }
\label{fig:Figure2}
\end{figure*}

The results of the detrending method for the $x$-coordinate of the chaotic Lorenz system can be seen in Fig. 1E-F. Here, $\sigma^2=20$ and 500 data points were used. It should be noted that with MVE, information from noisy time series belonging to the other two coordinates were also used in detrending of the noisy time series from the $x$-coordinate. In-sample cross-validation errors also correlate well with actual detrending errors (Fig. 1G and 1H). Optimized parameters from the in-sample cross-validation errors produces a noise-reduced time series that replicates the original deterministic time series remarkably well (Fig. 1F).  

Lastly, results of the detrending method for the chaotic Hindmarsh-Rose model can be seen in Fig. 1I-1H. Here, $\sigma^2=0.3$ and 500 data points were used. The Hindmarsh-Rose model is a model of neuronal spiking activity in the brain and is capable of chaotic behavior \cite{Hindmarsh1,WangPhysicaD}. The deterministic time series which we used consisted of a chaotic burst of spikes (Fig. 1J). From the in-sample cross-validation errors (Fig. 1K), there is difficulty in evaluating the performance of the forward algorithm and the bidirectional algorithm. Furthermore, the cross-validation errors do not correlate as well as the other two systems. These problems are, in this case, presumably due to the considerable noise involved since these problems alleviated with a smaller amount of observational noise (see Appendix). The parameters determined with the in-sample forecasting errors are the bidirectional algorithm with three recursive counts (Fig. 1K). These parameters are suboptimal according to the actual detrending errors (Fig. 1L). Nonetheless, even with suboptimal parameters, the noise-reduced time series still manages to resolve the spiking peaks rather well (Fig. 1J). Detrending errors and cross-validation errors for the other coordinates of the three systems analyzed also show that optimal or near-optimal parameters can be identified from the cross-validation errors (see Appendix). 

An obvious application of being able to detrend time series satisfactorily is to use the noise-reduced time series for the purposes of forecasting. By reducing the uncertainty in a training data set or library used to make forecasts, out-of-sample forecasts should be improved since there is less error in reconstructed state spaces \cite{KostelichE1}. In a similar vein, out-of-sample forecasts may also be used to determine the extent  of in-sample noise reduction. We made out-of-sample forecasts for the three systems analyzed in addition to a real-world data set on the pre-vaccination measles incidence rate from the state of New York which is at least partly chaotic due to the chaotic incidence rate of measles in New York City \cite{SugiharaNature1,Dalziel1,Schaffer1}. The noisy time series from Fig.1 and noisy time series for the other respective coordinates in the three systems were used as libraries. Out-of-sample forecasts one time step ahead with MVE (for the three multivariate systems) and simplex projection (for the measles data set) using the noisy time series were contrasted against those using the noise-reduced time series. Noise-reduced time series were obtained by objectively optimizing the heuristics based on the in-sample cross-validation errors (Fig. 1 and Appendix). In all systems, forecasts with noise-reduced time series produce less error than the noisy time series except the $y$-coordinate of the Hindmarsh-Rose model, which had a marginally higher error than the noisy time series (Fig. 2). This deviation of performance from the other systems and coordinates is notwithstanding the fact that using the dynamical trend as a library produces a better forecast (see Appendix), and also the fact that the detrending method did reduce the error of the time series as measured against $\mathbf{X}_t$ (Fig. S5). Therefore, a likely explanation for the poorer performance of the noise-reduced time series is that the detrending method had smoothened over certain sections of essential dynamics in the noisy time series. Nonetheless, this marginal decrease in forecasting performance in the $y$-coordinate should be measured against the more significant increase in forecasting performance in the $x$ and $z$-coordinate of the Hindmarsh-Rose model, of which the $x$-coordinate, the membrane potential, is the primary variable of interest in the model. These improved forecasts using the noise-reduced time series further demonstrates the ability of the detrending method to recover dynamics from noisy time series. 

While we have shown that the two heuristics introduced here can be optimized with in-sample cross-validation errors, it is conceivable that other parameters such as the embedding dimension (which we had set at 3 for this study), number of variable and lag combinations to use (for MVE), and $\alpha$ may also be optimized with the in-sample errors. The optimization of these parameters and other potential ones identified by Ye and Sugihara in MVE \cite{YeScience1} may provide room for greater improvement to the detrending performance of the detrending method. We refrain from exploring any of these other parameters in detail so as not to depart from the intention of this work as a concise presentation on a simple and multivariate non-parametric detrending technique. 

There exists strong theoretical justifications for the use of state space reconstruction as a non-parametric forecasting technique. This makes it ideal for its use in non-parametric detrending and grounds the detrending method in rigor and objectivity. Moreover, the results presented in this paper demonstrate the efficacy of the detrending method. Therefore, through its use in uncovering inherent dynamics from noisy time series, we envision that the detrending method introduced here will help shed new insights on the dynamics of complex systems which are not well understood, and for which no satisfactory model exists. 

\section*{Acknowledgements}
The author is grateful to Chew Lock Yue and Christopher Monterola for comments on the manuscript.

\renewcommand\thefigure{S\arabic{figure}}
\renewcommand\theequation{S\arabic{equation}}
\setcounter{figure}{0}

\section*{Appendix}
\subsection*{Van der Pol oscillator}
The dynamical equations for the Van der Pol oscillator are
\begin{align}
\dot{x} &= y,\\
\dot{y} &= \mu(1-{x}^2)y-x. 
\end{align}
The dynamical equations were integrated with an RK4 method and the step size of integration is 0.01. The initial condition for integration was (1, 1) and 600 time steps were discarded initially before sampling to allow the system to decay towards the limit cycle. Time series were then downsampled at a ratio of 5:1 to give $\mathbf{X}_t$ which consists of 800 points in each coordinate. A white noise vector $\bm{\epsilon}_t$ of variance 0.1 was added to $\mathbf{X}_t$ to give $\mathbf{Y}_t$. A total of  5 oscillations were used corresponding to the 800 data points in each coordinate of $\mathbf{X}_t$. The out-of-sample data set consists of 10,000 downsampled data points in each coordinate 2,081 downsampled points after the end of $\mathbf{X}_t$. 

\begin{figure*}[h!]
\centering
\includegraphics[scale=0.4]{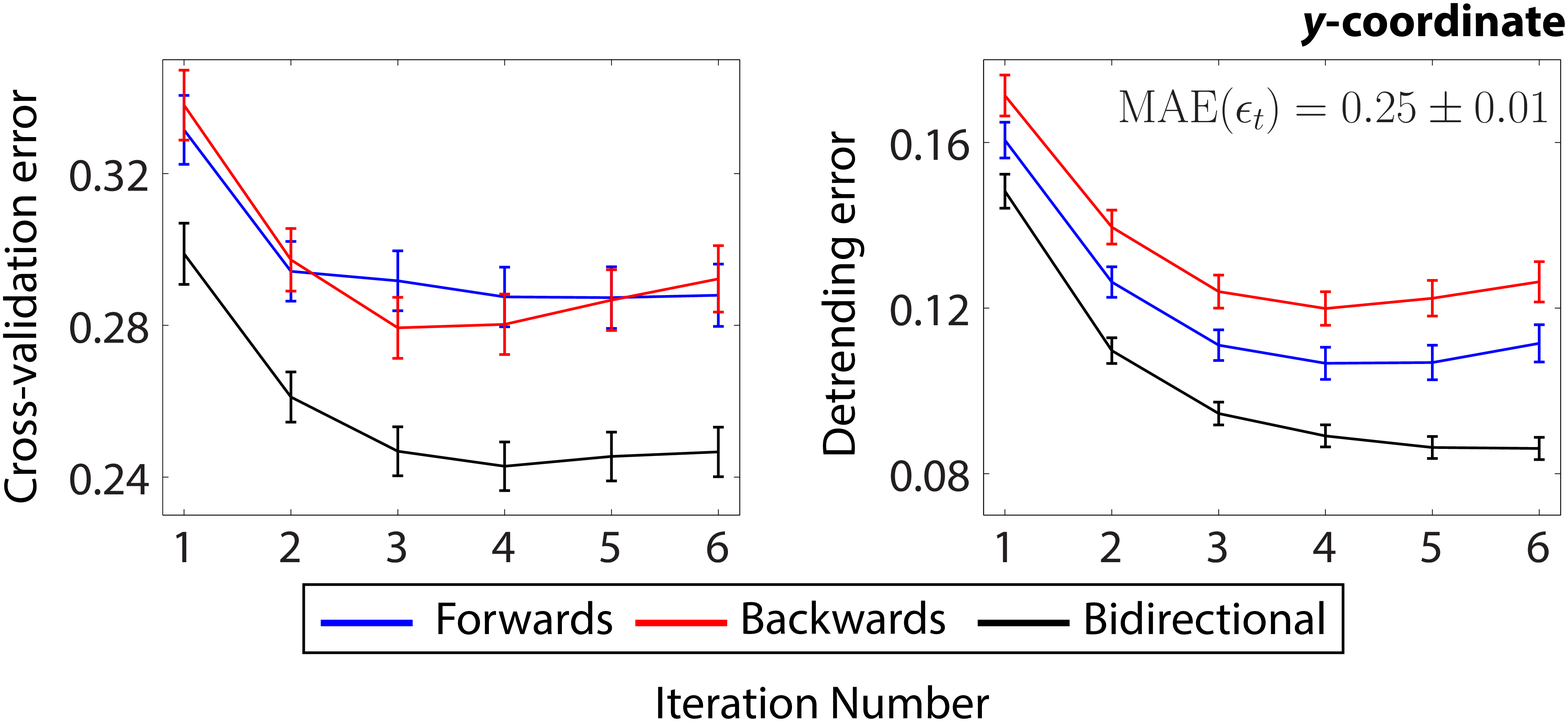}
\caption{The cross-validation error and actual detrending errors for the $y$-coordinate of the Van der Pol oscillator with $\sigma^2$=0.1. The optimal parameters as determined from the cross-validation error are the bidirectional algorithm with four recursive counts, close to the actual optimal parameters from the detrending errors. }
\label{fig:VDP_y}
\end{figure*}

\begin{figure*}[h!]
\centering
\includegraphics[scale=0.45]{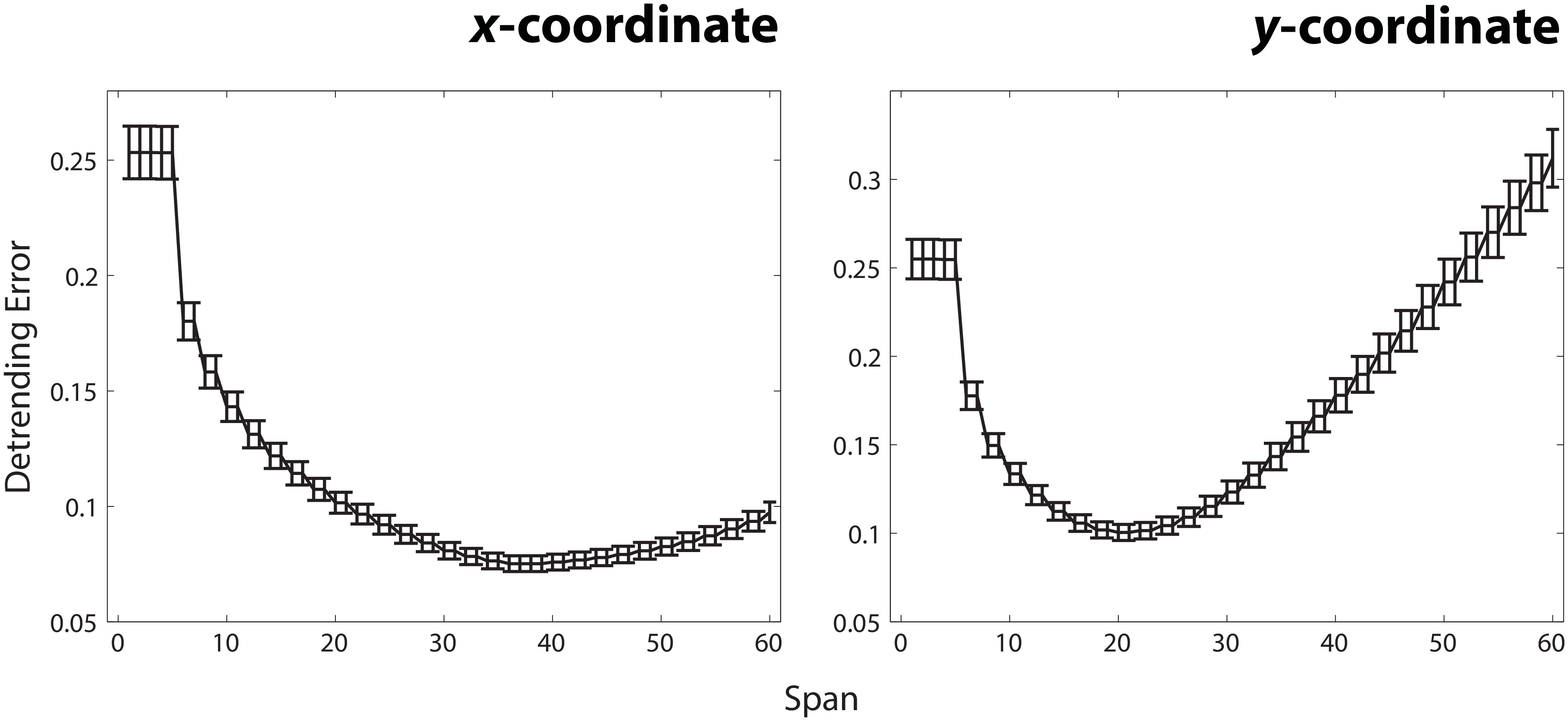}
\caption{The detrending error against span for the Loess smoothing on $\mathbf{Y}_t$ from the Van der Pol oscillator with $\sigma^2$=0.1. }
\label{fig:VDP_span}
\end{figure*}

$\mathbf{Y}_t$ was also corrected with Loess smoothing. This was accomplished with the \texttt{smooth} function in \texttt{MATLAB}. The detrending error vs span for both coordinates can be seen in Figure \ref{fig:VDP_span} for the time series corrected by Loess smoothing. 

The optimal span as calculated from the detrending errors is 36 for the $x$-coordinate and 20 for the $y$-coordinate. To combine detrending errors from both coordinates, we calculate the combined detrending error calculated which uses the norm instead to give a scalar output i.e. $\Expectation[||\mathbf{\hat{Z}}_t-\mathbf{X}_t||]$, where $\mathbf{\hat{Z}}_t$ is the corrected time series. The combined detrending error for the detrending method with the objectively optimized parameters (Figure 1 and Figure \ref{fig:VDP_y}) using five recursive iterations and the bidirectional algorithm for both coordinates is 0.1316$\pm$0.0028 whereas that from Loess smoothing using the optimal spans for both coordinates (Figure \ref{fig:VDP_span}) is 0.1393$\pm$0.0001. 

\subsection*{The Lorenz system}
The dynamical equations for the chaotic Lorenz system analyzed are
\begin{align}
\dot{x} &= 10(y-x), \\
\dot{y} &= x(28-z)-y, \\
\dot{z} &= xy-\frac{8}{3}z. 
\end{align}
\begin{figure*}[h!]
\centering
\includegraphics[scale=0.4]{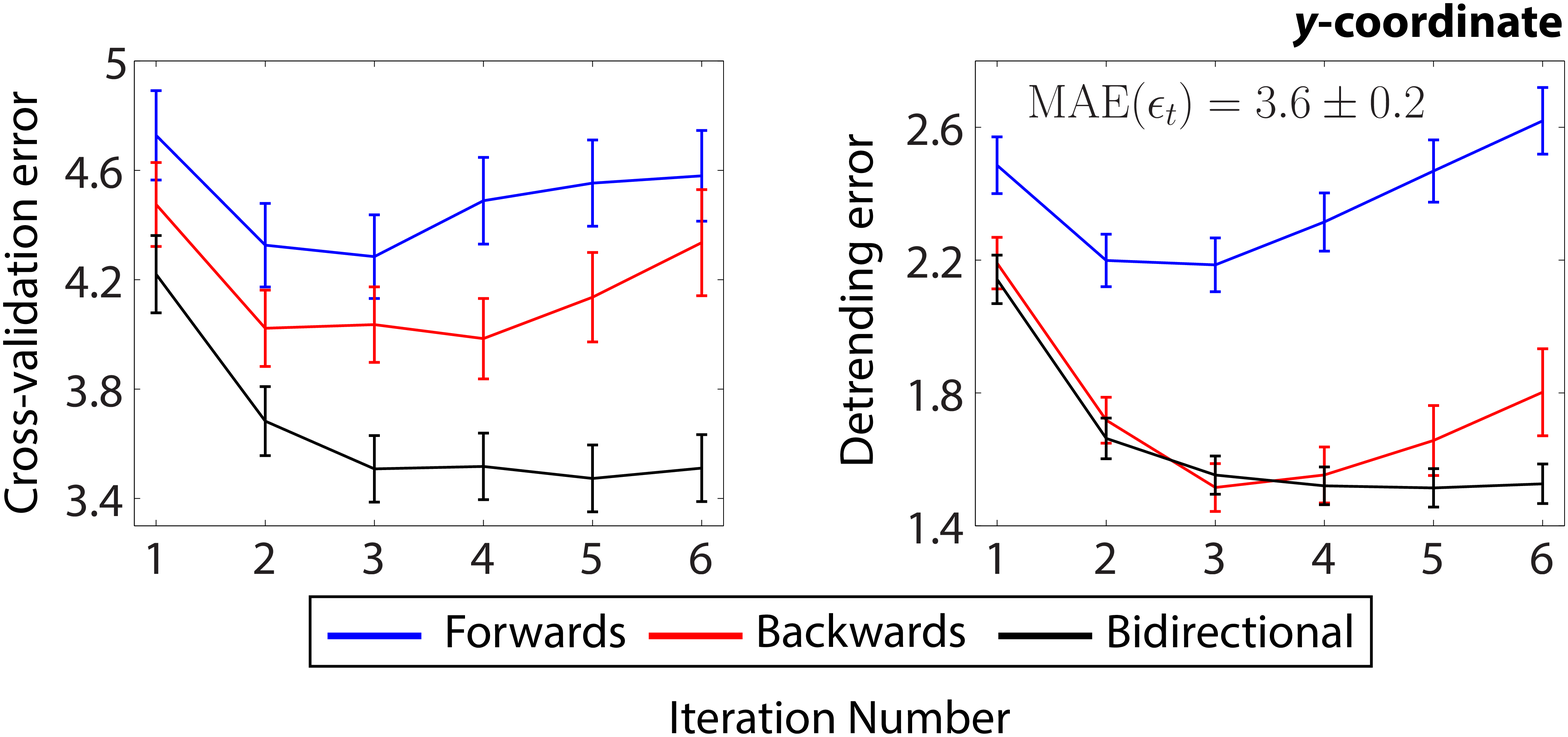}
\caption{The cross-validation error and actual detrending errors for the $y$-coordinate of the Lorenz system with $\sigma^2$=0.2. The optimal parameters as determined from the cross-validation error are the bidirectional algorithm with five recursive counts, which are the actual optimal parameters from the detrending errors. }
\label{fig:Lorenz_y}
\end{figure*}
\begin{figure*}[h!]
\centering
\includegraphics[scale=0.4]{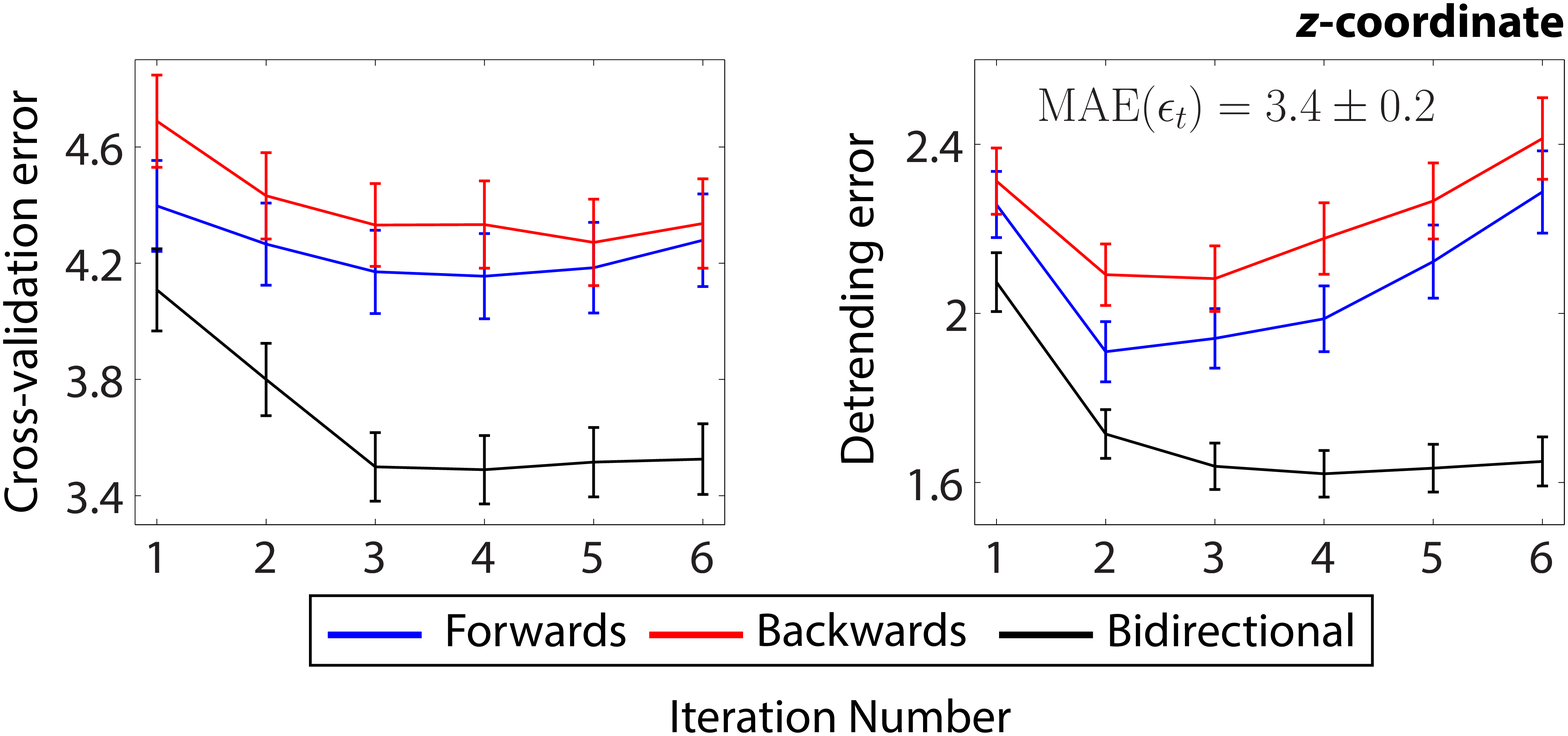}
\caption{The cross-validation error and actual detrending errors for the $z$-coordinate of the Lorenz system with $\sigma^2$=0.2. The optimal parameters as determined from the cross-validation error are the bidirectional algorithm with four recursive counts, which are the actual optimal parameters from the detrending errors. }
\label{fig:Lorenz_z}
\end{figure*}
The dynamical equations were integrated with an RK4 method and the step size of integration is 0.01. The initial condition for integration was (1, 1, 1) and 600 time steps were discarded initially before sampling to allow the system to decay towards chaotic attractor. Time series were then downsampled at a ratio of 5:1 to give $\mathbf{X}_t$ which consists of 500 data points in each coordinate. A white noise vector $\bm{\epsilon}_t$ of variance 20 was added to $\mathbf{X}_t$ to give $\mathbf{Y}_t$. The out-of-sample data set consists of 10,000 downsampled data points in each coordinate 2,381 downsampled points after the end of $\mathbf{X}_t$. 

\subsection*{The Hindmarsh-Rose model}
The dynamical equations for the chaotic Hindmarsh-Rose system analyzed are
\begin{align}
\dot{x} &= y-x^3+3x^2-z, \\
\dot{y} &= 1-5x^2-y, \\
\dot{z} &= 0.004[x-(z-3.19)/4].
\end{align}
\begin{figure*}[h!]
\centering
\includegraphics[scale=0.4]{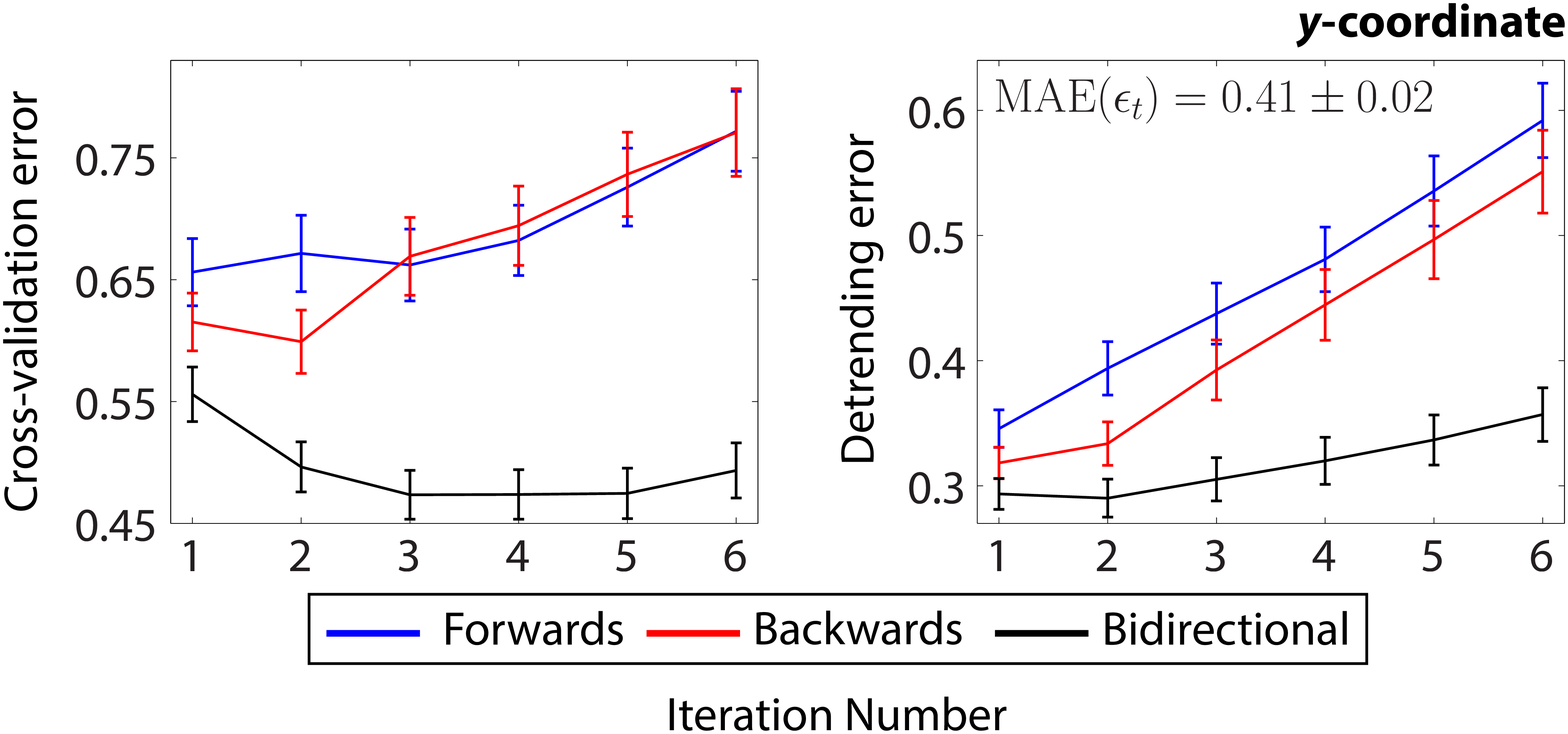}
\caption{The cross-validation error and actual detrending errors for the $y$-coordinate of the Hindmarsh-Rose system with $\sigma^2$=0.3. The optimal parameters as determined from the cross-validation error are the bidirectional algorithm with three recursive counts, close to the actual optimal parameters from the detrending errors. }
\label{fig:HindmarshRose_y}
\end{figure*}
\begin{figure*}[h!]
\centering
\includegraphics[scale=0.4]{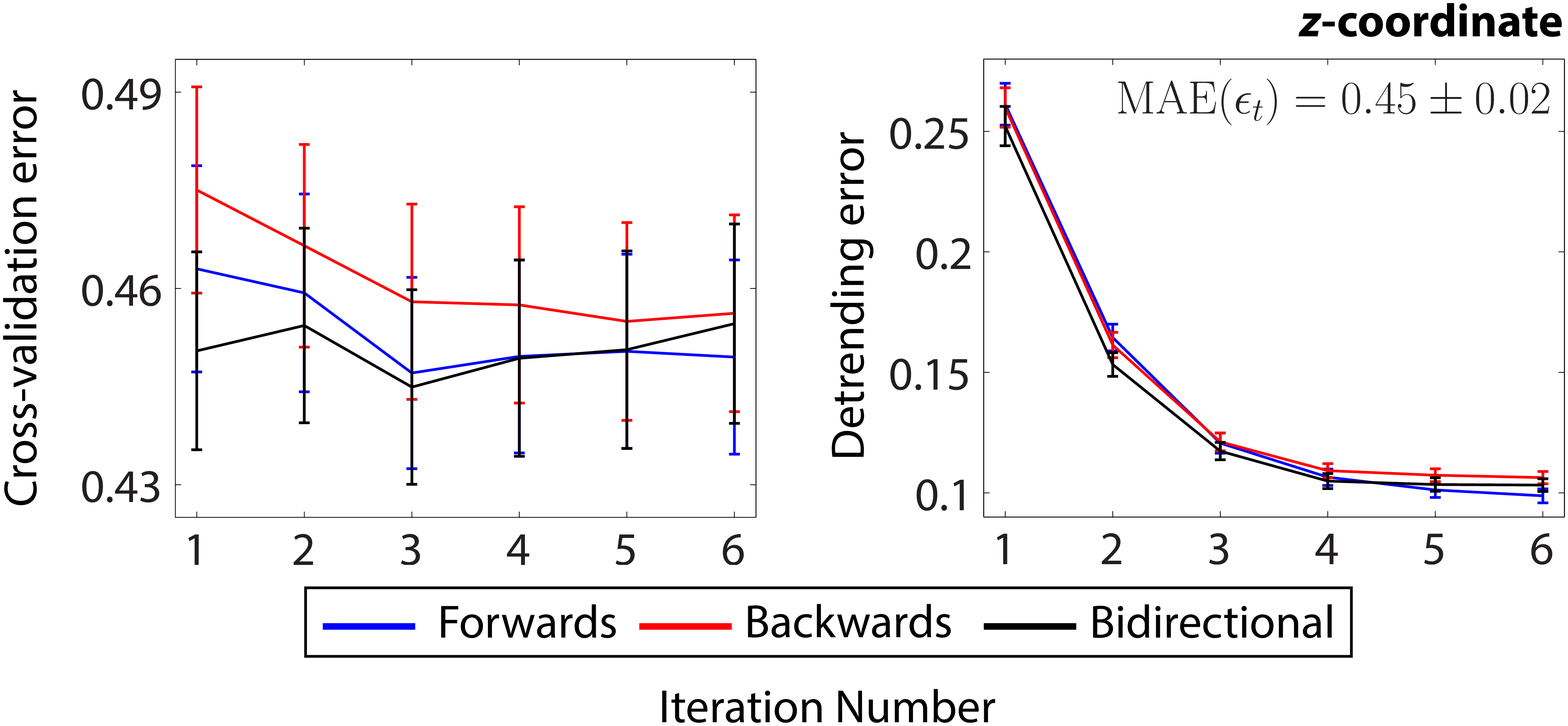}
\caption{The cross-validation error and actual detrending errors for the $z$-coordinate of the Hindmarsh-Rose system with $\sigma^2$=0.3. The optimal parameters as determined from the cross-validation error are the bidirectional algorithm with three recursive counts, close to the actual optimal parameters from the detrending errors. }
\label{fig:HindmarshRose_y}
\end{figure*}
\begin{figure*}[h!]
\centering
\includegraphics[scale=0.35]{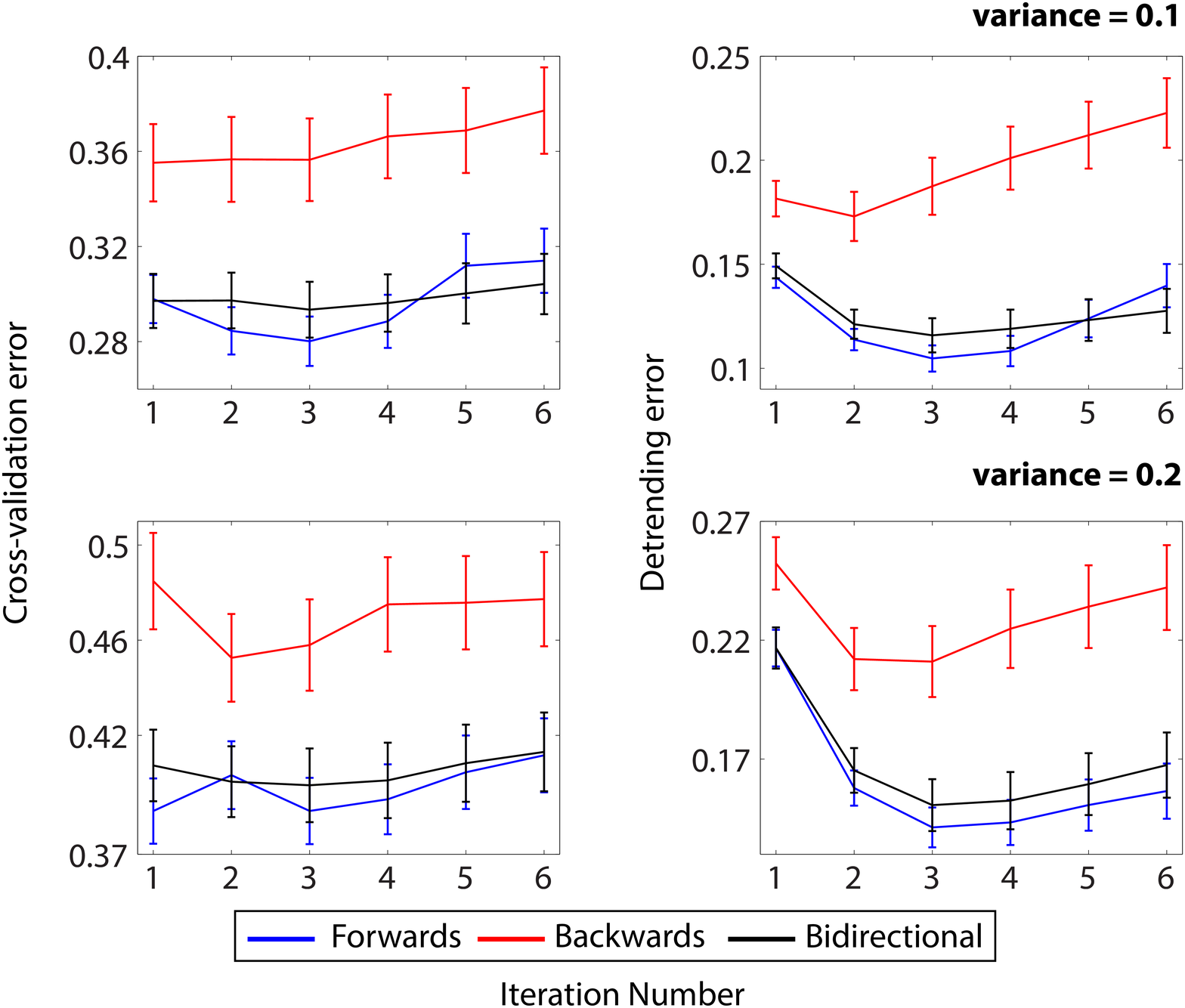}
\caption{In-sample cross-validation errors and detrending errors vs the recursive iteration count for white noise with lower variances in the Hindmarsh-Rose model analyzed. }
\label{fig:HindmarshRose_LowerNoise}
\end{figure*}

The dynamical equations were integrated with an RK4 method and the step size of integration is 0.2. The initial condition for integration was (1, 1, 1) and 100 time steps were discarded initially to allow the system to decay towards the attractor. Time series were then downsampled at a ratio of 5:1. Another 550 time steps were then discarded from this downsampled time series so that the new time series starts at the beginning of a chaotic bursts of spikes. This gives $\mathbf{X}_t$ which consists of 500 data points in each coordinate. A white noise vector $\bm{\epsilon}_t$ of variance 0.3 was added to $\mathbf{X}_t$ to give $\mathbf{Y}_t$. The out-of-sample data set consists of 10,000 downsampled data points in each coordinate 931 downsampled points after the end of $\mathbf{X}_t$. 

From Fig. \ref{fig:HindmarshRose_LowerNoise}, the in-sample performance errors of the bidirectional algorithm and the forward algorithm becomes easier to differentiate as compared to the higher noise used in the main text. Furthermore, the in-sample cross-validation errors correlate better with the actual detrending errors, yielding optimal parameters for the detrending method. 

The out-of-sample forecast performance of the dynamical trend as a library is $1.48\pm0.01$ (normalized MAE), lesser than the performance of the forecast by using the noisy time series as a library (Fig. 2). 

\subsection*{The measles data set}
The measles data set for the state of New York was obtained from the \href{https://www.tycho.pitt.edu/data/level1.php}{Project Tycho} database (accessed 31 Oct 2016). The data is the weekly incidence rate of measles per 100,000 population from 1928 onwards. The measles vaccine was introduced in 1963. Therefore, the data we used was truncated at the end of the last week of 1961. Missing data points were sparse and were interpolated with a cubic spline. The time series was then partitioned into 50-50 portions of an in-sample library and an out-of-sample time series. 
\begin{figure*}[h!]
\centering
\includegraphics[scale=0.4]{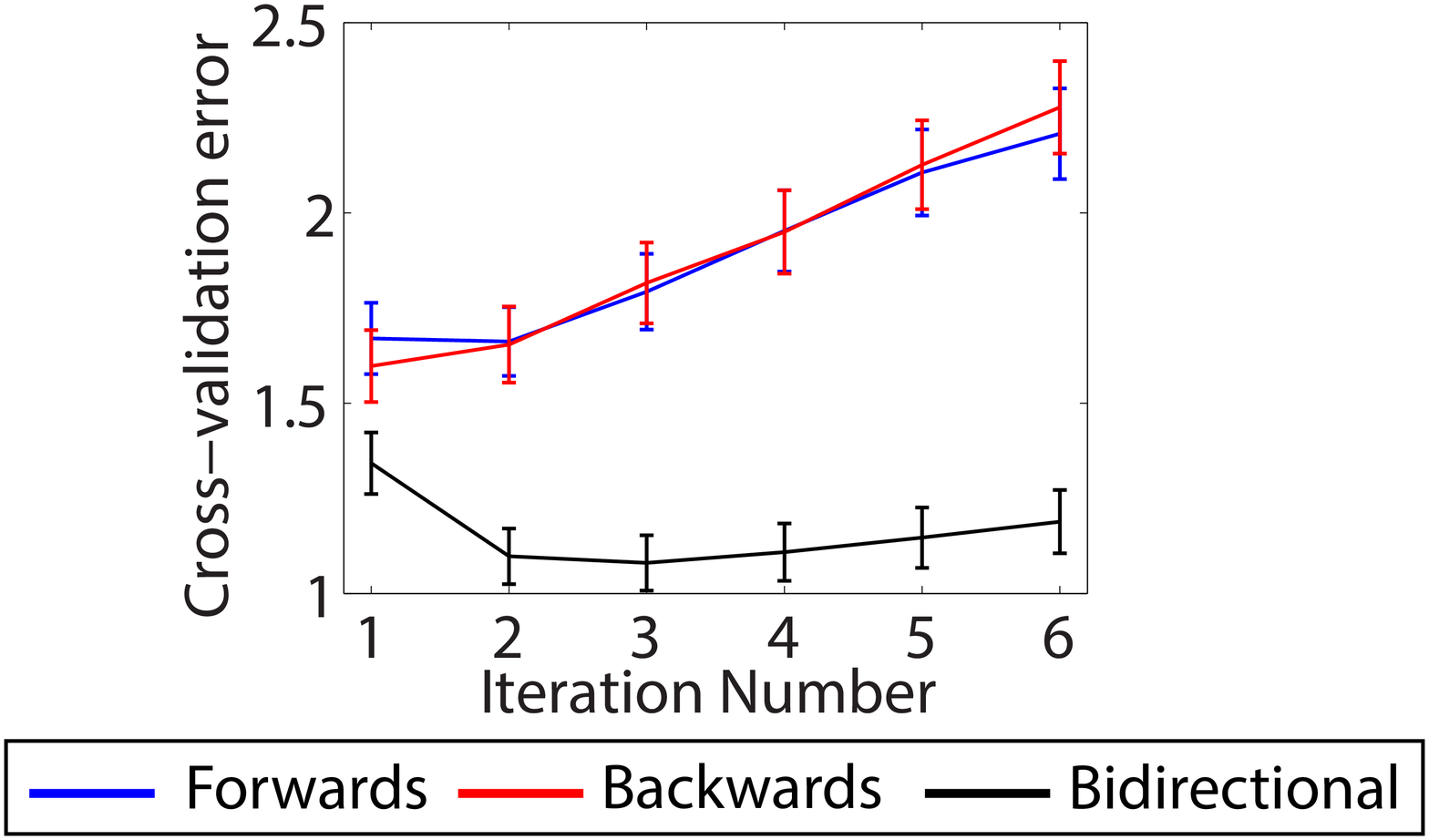}
\caption{In-sample cross-validation errors vs the recursive iteration count for the library of the measles data set.}
\label{fig:Measles_ErrorINSAM}
\end{figure*}

From the cross-validation errors (Fig. \ref{fig:Measles_ErrorINSAM}), we choose the bidirectional algorithm with three recursive iterations to detrend the time series.

\end{document}